\documentclass[aps,reprint,shownopacs,superscriptaddress]{revtex4-2}
\usepackage{graphicx}
\usepackage{amssymb,amsmath,amsthm,bm}
\usepackage{lipsum}
\usepackage{xcolor}
\usepackage{natbib}
\usepackage[colorlinks=true, linkcolor=blue,citecolor=red!75!black, urlcolor=red!75!black, pdfauthor={ },pdftitle={ },pdfsubject={ },pdfkeywords={ }]{hyperref}
\graphicspath{{Figures/}}

\usepackage[small,raggedright]{titlesec}
\titleformat*{\section} {\bf}
\titlespacing*{\section} {0pt}{15pt}{0.2pt}
\titleformat*{\subsection} {\bf}
\titlespacing*{\subsection} {0pt}{10pt}{0pt}

\begin{document}

\title{Parallel Quantum Gates via Scalable Subsystem-Optimized Robust Control}

\author{Xiaodong Yang}
\email{yangxd@szu.edu.cn}
\affiliation{Institute of Quantum Precision Measurement, State Key Laboratory of Radio Frequency Heterogeneous Integration, College of Physics and Optoelectronic Engineering, Shenzhen University, Shenzhen 518060, China}
\affiliation{Quantum Science Center of Guangdong-Hong Kong-Macao Greater Bay Area (Guangdong), Shenzhen 518045, China}

\author{Ran Liu}
\affiliation{Quantum Science Center of Guangdong-Hong Kong-Macao Greater Bay Area (Guangdong), Shenzhen 518045, China}

\author{Jun Li}
\email{lijunquantum@szu.edu.cn}
\affiliation{Institute of Quantum Precision Measurement, State Key Laboratory of Radio Frequency Heterogeneous Integration, College of Physics and Optoelectronic Engineering, Shenzhen University, Shenzhen 518060, China}
\affiliation{Quantum Science Center of Guangdong-Hong Kong-Macao Greater Bay Area (Guangdong), Shenzhen 518045, China}

\date{\today}

\begin{abstract}
Accurate and efficient implementation of parallel quantum gates is crucial for scalable quantum information processing. However, the unavoidable crosstalk between qubits in current noisy processors impedes the achievement of high gate fidelities and renders full Hilbert-space control optimization prohibitively difficult. Here, we overcome this challenge by reducing the full-system optimization to crosstalk-robust control over constant-sized subsystems, which dramatically reduces the computational cost. Our method effectively eliminates the leading-order gate operation deviations induced by crosstalk, thereby suppressing error rates. Within this framework, we construct analytical pulse solutions for parallel single-qubit gates and numerical pulses for parallel multi-qubit operations. We validate the proposed approach numerically across multiple platforms, including coupled nitrogen-vacancy centers, a nuclear-spin processor, and superconducting-qubit arrays with up to 200 qubits. As a result, the noise scaling is reduced from exponential to linear for parallel single-qubit gates, and an order-of-magnitude reduction is achieved for parallel multi-qubit gates. Moreover, our method does not require precise knowledge of crosstalk strengths and makes no assumption about the underlying qubit connectivity or lattice geometry, thereby establishing a scalable framework for parallel quantum control in large-scale quantum architectures.
\end{abstract}

\maketitle

\section*{INTRODUCTIOIN}
The circuit model for quantum computation comprise layers of quantum gates constructed from a universal set of fundamental operations \cite{nielsen2010quantum}. Executing quantum gates in parallel reduces operation time and circuit depth \cite{maslov2008quantum}, leading to significant efficiency gains in quantum algorithms \cite{cleve2000fast,fowler2004implementation} and quantum simulation \cite{nam2019low,PhysRevLett.120.110501}. Parallel quantum gates are also key components for system characterization \cite{erhard2019characterizing,PRXQuantum.2.040338} and quantum error correction \cite{aharonov1997fault,moore2001parallel}. 
While high-fidelity entangling operations with isolated qubit pairs were achieved early on, large-scale parallel gates have only recently been demonstrated on platforms such as neutral atom arrays \cite{PhysRevLett.123.170503,bluvstein2022quantum,evered2023high}, superconducting circuits \cite{PhysRevLett.119.180511,cao2023generation,chen2025efficient,fan2025calibrating}, and trapped ions \cite{figgatt2019parallel}, but typically exceeding the fault-tolerant threshold \cite{arute2019quantum,PhysRevLett.127.180501}.
 Thus, realizing high-quality parallel gates is increasingly vital for leveraging quantum advantages on near-term devices \cite{preskill2018quantum}.

The primary obstacle to achieving high-fidelity parallel operations is the unwanted interactions, often referred to as crosstalk \cite{sarovar2020detecting,parrado2021crosstalk}, between subsystems that each gate operates on \cite{google2023suppressing,PRXQuantum.3.020301}. 
From the hardware perspective, crosstalk interactions are often difficult to remove effectively. Only a few systems, such as neutral atom arrays \cite{ebadi2021quantum}, can arrange qubits into designated gate sites with negligible crosstalk between them. In contrast, systems like nuclear spin systems \cite{RevModPhys.76.1037} and nitrogen-vacancy centers \cite{pezzagna2021quantum} have fixed qubit interactions that cannot be eliminated. Other systems, such as superconducting circuits \cite{krantz2019quantum} and semiconductor quantum dots \cite{chatterjee2021semiconductor}, allow adjustable qubit interactions, but still exhibit residual crosstalk effects \cite{PhysRevLett.129.040502,PRXQuantum.3.020301,PhysRevApplied.12.054023,PhysRevB.97.045431}.
From the control perspective, crosstalk terms can  be incorporated into the Hamiltonian for designing parallel gates  across the entire system space, but this becomes intractable as the system dimension grows exponentially. 
Tensor network methods are well suited for large-scale one-dimensional problems, but in higher dimensions or for long-time dynamics their effectiveness is strongly constrained by rapid entanglement growth and computational complexity \cite{berezutskii2025tensor}.
Dynamical decoupling can mitigate system-environment interactions but is primarily suited for small-scale local gates \cite{PhysRevB.97.045431,PhysRevApplied.18.024068,niu2024multi}. Recent efforts have focused on deriving analytical control conditions to suppress crosstalk, but these methods are typically limited to specific types of crosstalk and small-scale systems \cite{PhysRevLett.131.210802,figueiredo2021engineering,PhysRevApplied.21.024016}.
Therefore, a practical method to design high-fidelity parallel gates on large-scale quantum processors is urgently needed.

In this work, we propose a scalable crosstalk-robust control method to design general parallel quantum gates. 
The entire system is divided into subsystems according to the product structure of the parallel gates, such that each subsystem targets a local gate, with the assumption of weak inter-subsystem interactions. Instead of optimizing parallel gates across the entire system, we reformulate the problem as a robust control problem for individual subsystems, where ``robust" refers to minimizing the effects of inter-subsystem crosstalk.
This leads to a substantial reduction in the problem's scale and computational cost.
The robust control pulses are analytically designed via inverse geometric optimization \cite{PhysRevLett.125.250403,yang2024quantum} for parallel single-qubit gates, and via a combination of quantum optimal control with the Van Loan integral technique \cite{Vanload,haas_engineering_2019,PhysRevApplied.21.034042} for parallel multi-qubit gates.
In addition, our method avoids complex procedures for simulating crosstalk effects  \cite{PhysRevLett.126.230502}.
We demonstrate the effectiveness of our method by achieving high-fidelity parallel gates  on various quantum platforms, including parallel single-qubit gates on coupled NV centers with up to 9 spins, parallel single-qubit gates and parallel CNOTs on a 12-qubit nuclear spin processor, parallel CZ gates on a superconducting array up to 200 qubits. More importantly, our subsystem-optimized robust control method dramatically reduces the noise scaling associated with the number of parallel gates--bringing it from exponential to linear for single-qubit operations and achieving a 13-fold reduction for parallel CNOT gates.
The proposed method is valuable not only for quantum computation  but also for large-scale quantum simulation and metrology on near-term qubit platforms.

\begin{figure}
\includegraphics[width=0.92\linewidth]{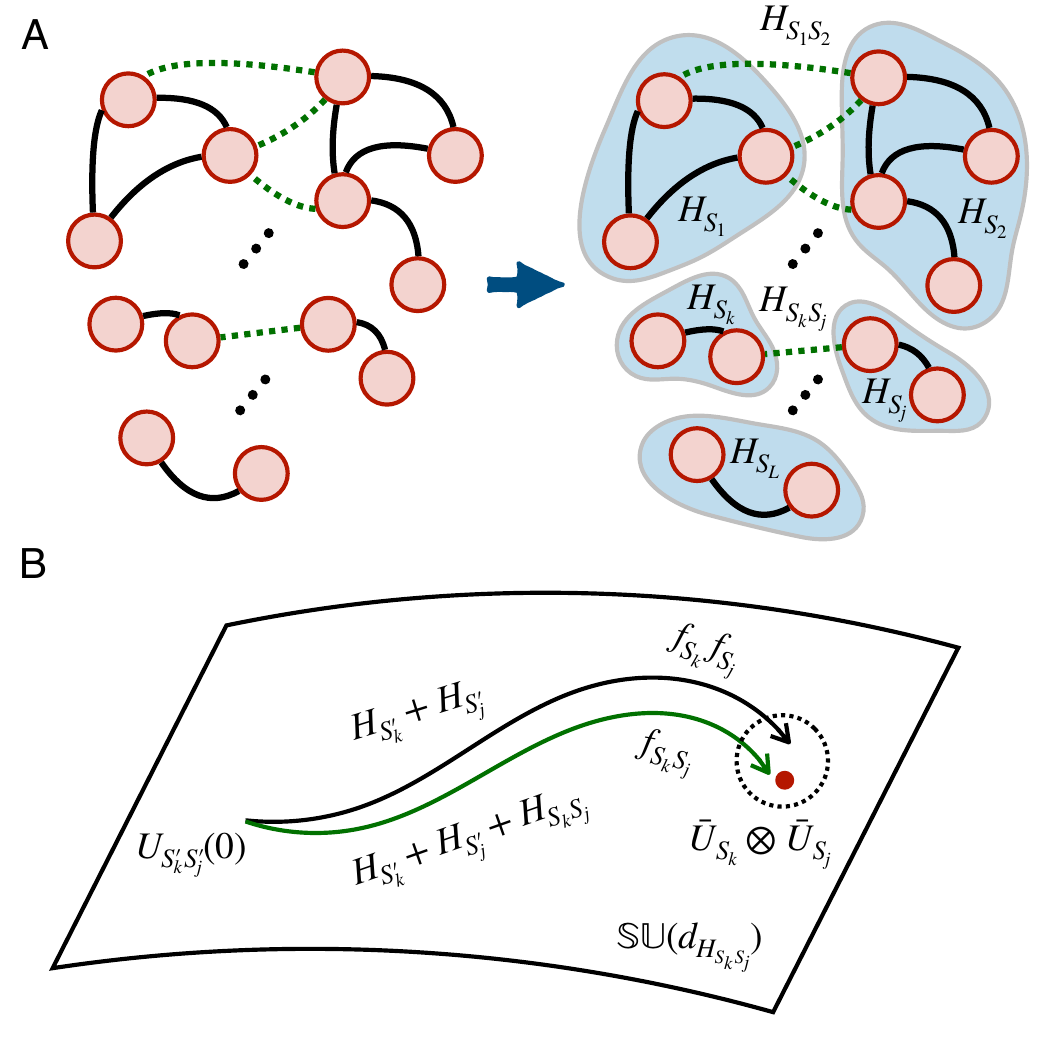}
\caption{Schematic diagram of the subsystem-optimized robust control for designing parallel quantum gates.
(a) Physically, it is reasonable to assume that each quantum gate acts on a single subsystem, with weak crosstalk between subsystems. The system can thus be divided into $L$ subsystems (blue shaded), each governed by a Hamiltonian $H_{S_k}$, with inter-subsystem crosstalk described by $H_{S_kS_j}$. 
(b) Subsystem-optimized robust control performs pulse optimization on each subsystem pair $S_k$ and $S_j$ for realizing $\bar U_{S_k}\otimes \bar U_{S_j}$, rather than the full system. Specifically, it maximizes the crosstalk-free fidelity $f_{S_k}f_{S_j}$ while minimizing the directional derivative $f_{S_k S_j}$ associated with the variation in $H_{S'_k}+H_{S'_j}$ along the direction $H_{S_k S_j}$.
  }
\label{subsystems}
\end{figure} 

\section*{RESULTS}
\subsection*{Problem formulation }
Consider an $N$-qubit quantum system governed by the system Hamiltonian $H_S$. The objective is to realize a specified yet widely used unitary operation, $\bar{U}$, composed as a tensor product of $L$ parallel quantum gates. To achieve the target gate, suitable control schemes governed by the control Hamiltonian $H_C$ must be designed, with system evolution described by the equation $U(t) = -i[H_S + H_C(t)]U(t)$, and the time evolution calculated as $	U(t) = \mathcal{T} \exp(  -i \int_0^t dt_1 \left[  H_S + H_C(t_1)   \right] )$, where $\mathcal{T}$ is the  time-ordered operator. The gate fidelity is typically assessed by 
\begin{equation}\label{gf}
	F = \left| \operatorname{Tr}(U(T) \overline{U}^\dagger) \right|^2 / d^2,
\end{equation}
 where  $T$ denotes the control period and $d=2^N$. 
Conventionally, this problem is addressed  using quantum optimal control, where optimal controls are determined by simulating the  system's evolution on classical computers. However, directly solving this problem is intractable due to the exponential growth in system dimensions.

\subsection*{Subsystem-optimized robust control}
Physically, the following two considerations are typically well-justified: (i) each local gate is associated with a specific subsystem, allowing the target operation to be written as $\bar U= \bar U_{S_1}\otimes \cdots \otimes \bar U_{S_L}$. (ii) there are only a few comparatively small crosstalk couplings between the subsystems. For example, in superconducting-qubit systems, unwanted couplings can be largely turned off during parallel gate operations, but small and unavoidable crosstalk interactions between subsystems still remain. Therefore, we can divide the entire system into $L$ disjoint subsystems: $S=S_1 \cup \cdots \cup S_L$.  The division may vary for different parallel gates within a quantum circuit.
Under this division, the system Hamiltonian is expressed as 
\begin{equation}\label{sysham}
	H_S=H_0+H_1=\sum_{k=1}^L H_{S_k}+ \sum_{k<j}^L H_{S_k S_j},
\end{equation}
where $H_{S_k}$ is the internal Hamiltonian of subsystem $S_k$, and $H_{S_k S_j}$ represents the crosstalk Hamiltonian between subsystem $S_k$ and $S_j$.
 We further assume that  $H_{C_k}(t)$ denotes the control Hamiltonian acting on subsystem $S_k$, and we define $H_{S'_k}\equiv H_{S_k}+H_{C_k}(t), H'_0 \equiv \sum_{k=1}^L H_{S'_k}$.


The core idea of our approach is to frame the problem as a robust quantum control task, solving the control fields for subsystems while ensuring robustness against the crosstalk couplings $H_1$.
Specifically, we treat $H_{1}$ as a perturbation, and expand the time-evolution operator using the Dyson series \cite{PhysRev.75.486} as
$
 U(T)=U_{H_0'}(T)  [ \mathbb{I}_d  -i\int_0^T dt U_{H_0'}^\dag(t)H_1 U_{H_0'}(t)- \int_0^T d t_1 \int_0^{t_1} d t_2 U_{H_0'}^\dag(t_1)H_1 U_{H_0'}(t_1)U_{H_0'}^\dag(t_2)H_1 U_{H_0'}(t_2) +\cdots ]
$
where $\mathbb{I}_d$ is the $d$-dimensional identity matrix, 
$U_{H_0'}(t) = \bigotimes_{k=1}^L  U_{S_k'}(t)$ represents the crosstalk-free evolution operator.
This conveniently separates the crosstalk effects from the overall evolution.
Furthermore, by truncating the above series to first- and second-order terms, the gate fidelity in Eq. (\ref{gf}) approximates to
\begin{equation}\label{DysonSeries}
	F\approx 1-\frac{1}{d} \sum_{k<j}^L \|\mathcal{D}_{U_{H_0'}(T)}(H_{S_k S_j})\|^2,
\end{equation}
where 
\begin{equation*}
\mathcal{D}_{U_{H_0'}}(H_{S_k S_j})=-i U_{H_0'}(T)\int_0^T dt U_{H_0'}^\dag(t)H_{S_k S_j} U_{H_0'}(t)
\end{equation*}
represents the first order directional  deviation of $U_{H_0'}(T)$   induced by a variation in $H'_0$ along the direction $H_{S_k S_j}$ \cite{haas_engineering_2019} and  $\| \cdot \|$ denotes the Frobenius norm.  The problem now reduces to finding controls that maximize the crosstalk-free gate fidelity $f_{0} = \left| \operatorname{Tr}(U_{H_0'}(T) \overline{U}^\dag) \right|^2 / d^2 $ while minimizing the norm of each directional derivative $f_{S_k S_j} \equiv \|\mathcal{D}_{U_{H_0'}(T)}(H_{S_k S_j})\|^2 / d $; see details in Methods and Supplemental Material.

A key observation is that, to obtain each term $f_{S_k S_j}$, it suffices to perform computations solely within subsystems $S_k$ and $S_j$. Specifically, we notice that the directional deviation in Eq. (\ref{DysonSeries})  can be simplified to
\begin{equation}
 	\mathcal{D}_{U_{H_0'}(T)}(H_{S_k S_j})= \mathcal{D}_{U_{S'_k S'_j}(T)} (H_{S_k S_j})  \otimes U_{S/ \left\{S_k, S_j \right\}}(T), \nonumber
 \end{equation}
where we denote $U_{S'_kS'_j}(T)=U_{S_k'}(T) \otimes U_{S_j'}(T)$ and $U_{S/ \left\{S_k, S_j \right\}}(T) = \bigotimes^L_{l \ne k,j} U_{S_l'}(T)$.
This implies that each $f_{S_k S_j}$ can be evaluated independently on a pair of subsystems, avoiding the need for computations over the entire system space, i.e.,
\begin{equation}\label{ddnorm}
	f_{S_k S_j} = \|\mathcal{D}_{U_{S'_kS'_j}(T)}(H_{S_k S_j})\|^2 / d_{H_{S_k S_j}}.
\end{equation}
Moreover, the crosstalk-free gate fidelity can be independently evaluated within each subsystem as $f_0=\prod_{k=1}^L f_{S_k}$ and $f_{S_k}=\left| \operatorname{Tr}(U_{S'_k}(T) \overline{U}^\dag_{S_k}) \right|^2 / d_{H_{S_k}}^2 $.
Therefore, the robust control problem  essentially becomes a multi-objective optimization problem on subsystems
\begin{align}
\label{QOC}
	\max \quad  &  f = \prod_{k=1}^L f_{S_k} - \sum_{k<j}^L \lambda_{kj} f_{S_k S_j} ,  \\
    \text{s.t.} \quad & \dot U_{S'_k}(t) = -i[H_{S_k} +H_{C_k}(t)]U_{S'_k}(t), \nonumber
\end{align}
where $\lambda_{kj}$ represents positive weight coefficients requiring careful adjustment during optimization. Details are provided in Supplemental Material.
 It is worth noting that the number of directional derivatives $f_{S_k S_j}$ is at most  $L(L-1)/2$, and each subsystem typically contains only a few qubits.  Consequently, the proposed subsystem-optimized robust control is generally efficient for designing large-scale parallel quantum gates.

\begin{figure*}
\includegraphics[width=0.9\linewidth]{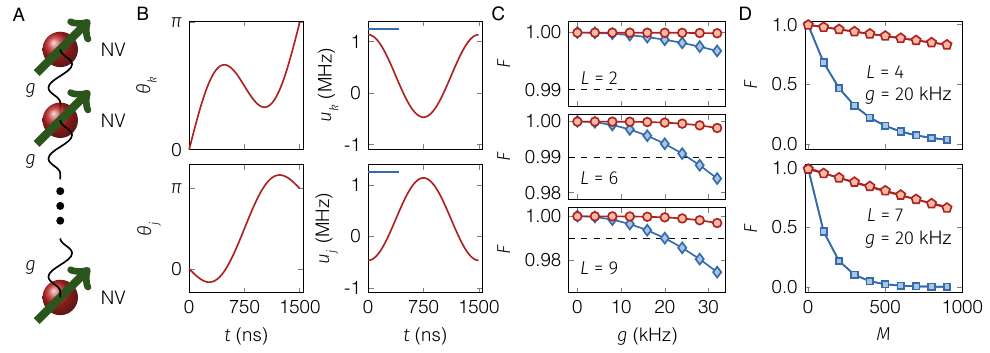}
\caption{\textbf{Implementation of parallel single-qubit $R_y(\pi)$ gates on  coupled NV centers.} (\textbf A) Dipolar coupled NV centers with homogeneous coupling strengths $g$.  (\textbf B) Geometric trajectories $\theta_k, \theta_j$ and the corresponding robust control fields $u_k, u_j$ for each subsystem pair $(k,j)$. For $N$ coupled NV centers, $u_k$ and $u_j$ are applied in an alternating manner to the electron spins. For comparison, the primitive rectangular pulse $u_p$ (blue lines) is also demonstrated, which is applied to each electron spin identically. (\textbf C) Gate fidelities $F$ versus the coupling strength $g$ for different number of subsystems $L$. The dashed lines indicate the fidelity threshold of 0.99. (\textbf D) Gate fidelities $F$ as a function of the number of repeated parallel gates $M$. The pentagon markers denote the linear-fit results, while the square markers represent the exponential-fit results.  
} \label{NVChain}
\end{figure*}

\subsection*{Parallel single-qubit quantum gates}
To design high-fidelity parallel single-qubit gates under inter-subsystem crosstalk, we employ inverse geometric optimization \cite{PhysRevLett.125.250403,yang2024quantum}. For clarity, we consider the case where each subsystem consists of a single qubit. 
Each qubit is typically driven by a local control Hamiltonian of the form  $H_{C_k} (t)=u_k (t) [\cos \phi_k(t) \sigma_x^k/2 + \sin \phi_k(t)\sigma_y^k /2]$, where $\sigma_{x(y)}^{k}$ are Pauli-X(Y) operators acting on the $k$th qubit, $u_k(t)$ and $\phi_k(t)$ are the instantaneous amplitude and phase of the driving field, respectively. 
We further assume a typical crosstalk  between subsystems $S_k$ and $S_j$, given by $H_{S_k S_j}=g_{kj} \sigma_{z}^k \otimes \sigma_{z}^j$, where $\sigma_z^{k(j)}$ are Pauli-Z operators acting on the $k$th ($j$th) qubit, respectively, and $g_{kj}$ represents the  coupling strength.  The  norm of each directional deviation $f_{S_k S_j}$ in Eq. (\ref{ddnorm}) can then be expressed as
\begin{align}\label{fsksj}
	f_{S_k S_j} &=\frac{1}{4} \left \|\int_0^T U_{S'_kS'_j}^\dag(t)  H_{S_k S_j} U_{S'_kS'_j}(t)  d t \right\|^2 \nonumber \\
	&= \frac{g_{kj}^2}{4} \sum\limits_{\mu,v=x,y,z} \left( \int_0^T r_\mu^k (t)  r_v^j (t) d t\right)^2,
\end{align}
where $r_\mu^k (t)= \text{Tr}[U_{S'_k}^\dag(t) \sigma_{z}^k  U_{S'_k}(t) \sigma_\mu^k]/2$ and $r_v^j (t)= \text{Tr}[U_{S'_j}^\dag(t) \sigma_{z}^j  U_{S'_j}(t) \sigma_v^j]/2$; see details in Methods. 
	 
Subsequently, we geometrically parameterize the crosstalk-free  evolution of each subsystem as  $U_{S'_k}(t)= \exp(i\beta) R_z(\varphi_k)R_y(\theta_k)R_z(\gamma_k)$, where $\beta, \theta_k, \varphi_k,\gamma_k \in [-\pi,\pi]$. Here, we further set $\beta=0$ because $H_{S'_k}$ is traceless. This decomposition enables direct correspondence between the unitary trajectory and the control parameters.
As such, we obtain $r_x^k(t)=-\cos\gamma_k \sin \theta_k, r_y^k(t)=\sin\gamma_k \sin \theta_k, r_z^k(t)=\cos \theta_k$ and similarly  for $r_v^j(t)$. 
The design of robust pulses requires $f_{S_k S_j}=0$, which further implies $\int_0^T r_\mu^k (t)  r_v^j (t) d t=0$ for all $\mu,v=x,y,z$. Additional constraints  arise from the choice of the specific quantum gate, as well as from the Schr{\"o}dinger equations  (derived from  Eq. (\ref{QOC})):  $\dot \theta_k  = u_k  \sin(\phi_k  - \varphi_k ), \dot \varphi_k   =  -u_k \cos(\phi_k  - \varphi_k )   \cot \theta_k, \dot \gamma_k     = u_k \cos(\phi_k  - \varphi_k ) /\sin  \theta_k $. 

\subsection*{Parallel multi-qubit quantum gates}
For parallel multi-qubit quantum gates, geometric optimization is less practical. The main difficulty lies in efficiently computing the directional deviations, which we address using the Van Loan integral technique \cite{Vanload,haas_engineering_2019,PhysRevApplied.21.034042}.
Concretely, we define a set of operators $\left\{L_{S_k S_j}: 1\le k < j \le L \right\}$ and $\left\{ L_{C_k}(t): k =1,\ldots, L\right\}$ as follows 
\begin{align}
	L_{S_k S_j} & = \left[\begin{matrix}
H_{S_k} + H_{S_j}    &  H_{S_k S_j} \\
0  & H_{S_k} + H_{S_j}   \\
\end{matrix}
\right],  \\
L_{C_k }(t) & = \left[\begin{matrix}
H_{C_k}(t)     &  0 \\
0  & H_{C_k}(t)   \\
\end{matrix}
\right]. \nonumber
\end{align}
Assuming they satisfy the differential equation $ \dot V_{S'_k S'_j}(t)  =   -i  [L_{S_k S_j}    +  L_{C_k}(t)  + L_{C_j}(t) ] V_{S'_k S'_j}(t)$,  its solution can be expressed as  
\begin{equation}
	V_{S'_k S'_j}(t) = \left[\begin{matrix}
U_{S'_k S'_j}(t)     &  \mathcal{D}_{U_{S'_k S'_j}(t)} (H_{S_k S_j}) \\
0  & U_{S'_k S'_j}(t)   \\
\end{matrix}
\right]. 
\end{equation}  
In this way, we can conveniently obtain each directional deviation from the off-diagonal element of the operator $V_{S'_k S'_j}$.
To further search for the robust control fields, various optimization algorithms can be employed. Here, we employ GRAPE \cite{khaneja2005optimal}, known for its excellent convergence speed and numerical accuracy. More details about the derivations and the algorithmic procedure can be found in the Supplementary Material.

\subsection*{Parallel gates on  coupled NV centers}
We first verify the proposed method by designing parallel single-qubit $R_y(\pi)$ gates on $N$ coupled NV centers \cite{dolde2013room,pezzagna2021quantum,PhysRevX.15.021069}. For simplicity, we consider neighboring dipolar-coupled NV centers with a homogeneous coupling strength, as depicted in Fig. \ref{NVChain}A. Typically, the dipole–dipole interaction is much smaller than the electron transition frequencies; therefore, the system Hamiltonian in the rotating frame can be effectively written as $H_S= g \sum_{k=1}^{N-1} \sigma_z^k  \sigma_z^{k+1}$, where $g$ represents the coupling strength. Each electron spin is driven by a field of the form $H_{C_k} (t)=u_k (t) [\cos \phi_k(t) \sigma_x^k/2 + \sin \phi_k(t)\sigma_y^k /2]$. 

To apply our subsystem-optimized robust control method, we partition the system into $L$ subsystems, each containing a single spin, so that $L=N$. 
Furthermore, we parameterize the crosstalk-free evolution of each subsystem as $U_{S'_k}(t)=R_y(\theta_k)$ (setting $\varphi_k=\gamma_k=0$ for simplicity) and apply inverse geometric optimization. For each pair of subsystems, the target quantum gate can be expressed as $e^{-i \pi \sigma_y^k /2} \otimes e^{-i \pi \sigma_y^j/2}$. At $t=0$, we set $U_{S'_kS'_j}(0)=I $, this corresponds to to the initial conditions $\theta_k(0)=\theta_j(0)=0$. The ending point conditions are $\theta_k(T)=\theta_j(T)=\pi$. 
To satisfy the above constraints together with $f_{S_k S_j}=0$, the analytical geometric evolution trajectories are obtained as
\begin{equation}
\begin{aligned}
\theta_k(t)&=\frac{\pi t}{T}+\frac{A}{2}\sin\left(\frac{2\pi t}{T} \right),  \\
\theta_j(t)&=\frac{\pi t}{T}-\frac{A}{2}\sin\left(\frac{2\pi t}{T} \right),
\end{aligned}
\end{equation}
where $A$ is the first positive zero of the Bessel function, i.e., $A\approx 2.404826$; see details in Supplemental Material. The corresponding control fields have the explicit form of $u_k(t)=\dot \theta_k$ and $u_j(t)=\dot \theta_j$. 

\begin{figure*}
\includegraphics[width=0.95\linewidth]{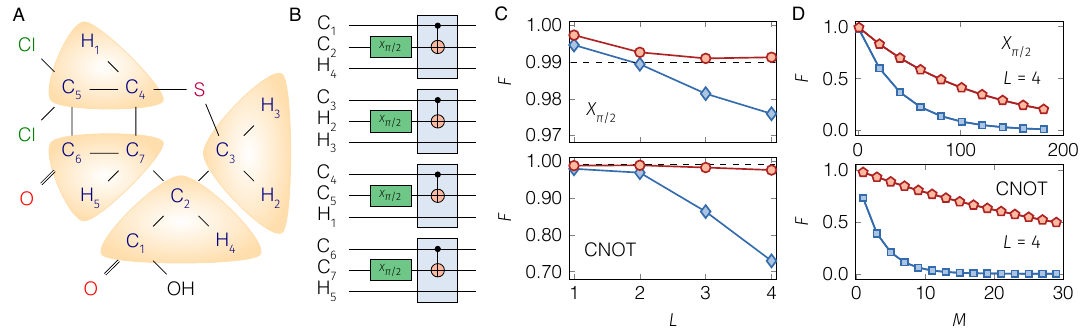}
\caption{\textbf{Implementation of parallel single-qubit $R_x(\pi/2)$  and parallel CNOT gates on  a 12-qubit NMR processor.} (\textbf A) Molecular structure of the NMR sample, divided into four yellow-shaded subsystems, each containing three spins. (\textbf B) Representative quantum circuit with parallel single-qubit and parallel CNOT gates acting on designated spins. (\textbf C) Gate fidelities $F$ versus the number of subsystems $L$, comparing our robust pulse (red lines) with the primitive pulse (blue lines) that does not account for crosstalk between subsystems. The dashed lines indicate the fidelity threshold of 0.99. (\textbf D) Gate fidelities $F$ versus the number of repeated parallel gates $M$. Both of the pentagon and square markers denote the exponential-fit results.} \label{NMR}
\end{figure*}

The geometric trajectories for each subsystem pair are shown in Fig.~\ref{NVChain}B, along with the corresponding control fields. The robust controls exhibit smaller maximum amplitudes to the primitive rectangular pulse $u_p$, while being longer in duration. We compare the gate fidelities of our robust pulse with those of the primitive pulse for $L=2,6,9$, as shown in Fig. \ref{NVChain}C. The results clearly show that our robust pulse demonstrates enhanced resilience to variations in coupling strength, achieving fidelities of above 0.99 across the tested range. Furthermore, we compare the gate fidelities obtained by repeatedly applying the parallel gates $M$ times for a fixed coupling $g$; see Fig.~\ref{NVChain}D. We observe that the primitive pulse exhibits an exponential fidelity decay, which is expected: if each gate carries a small infidelity $\varepsilon$, then repeating it $M$ times gives $F = (1-\varepsilon)^M \approx e^{-\varepsilon M}$. In contrast, our robust pulse shows a linear decay with $M$. The underlying reason is that the dominant lowest-order errors are strongly suppressed with our robust pulses, making $\varepsilon$ sufficiently small so that the exponential decay $e^{-\varepsilon M}$ is well approximated by its linear expansion $1-\varepsilon M$.
This demonstrates that the subsystem-optimized robust control reduces the noise scaling in parallel single-qubit gates from exponential to linear, a substantial improvement for scalable quantum computation.
 Therefore, these results confirm that subsystem-optimized robust control, combined with geometric optimization, is capable of  designing high-fidelity parallel single-qubit gates in a scalable manner.

\subsection*{Parallel gates on a  nuclear spin processor}
We then validate the developed method using a 12-qubit NMR processor \cite{PhysRevLett.123.030502,lu2017enhancing}; see  Fig. \ref{NMR}A.  Our targets include parallel single-qubit $R_x(\pi/2)$ gates and parallel CNOT gates, as illustrated in Fig. \ref{NMR}(B), which are essential components of many quantum circuits.
In the rotating frame, the system Hamiltonian takes the form  $H_S =  \sum_{i=1}^{12} {  \Omega_i   \sigma_z^i/2} + \pi\sum_{i<j}^{12} {J_{ij} \sigma_z^i \otimes \sigma_z^j/2}$, where $\Omega_i$ denotes the precession frequency of the $i$th spin, and $J_{ij}$ represents the coupling strength between spins $i$ and $j$. We assume individual controls $u(t)=(u_x^i(t),u_y^i(t)) ~(0\leq t \leq T)$, and the corresponding  Hamiltonian  can be expressed as $H_C(t) = \sum_{i=1}^{12} \left( u_x^i(t) \sigma_x^i + u_y^i(t) \sigma_y^i \right)$. 
This is to date the largest spin  quantum information processor with full controllability;  see Supplemental Materials for its Hamiltonian parameters. However, it remains challenging to design control pulses for implementing nontrivial quantum operations among the spins, since natural interactions exist between nearly all spin pairs and cannot be turned off. This typically requires computations involving matrix exponentials and multiplications in a $2^{12}$-dimensional Hilbert space, which is an intractable task on a standard personal computer.

We address this issue by first partitioning the entire system into four subsystems $S_1 = \left\{ \text{C}_1,\text{C}_2,\text{H}_4 \right\}$, $S_2 = \left\{ \text{C}_3,\text{H}_2,\text{H}_3 \right\}$, $S_2 = \left\{ \text{C}_4,\text{C}_5,\text{H}_1 \right\}$, and $S_4 = \left\{ \text{C}_6,\text{C}_7,\text{H}_5 \right\}$, each consisting of three spins; see Fig. \ref{NMR}A. Subsequently, we treat the spin interactions between different subsystems as crosstalk and apply our subsystem-optimized control method.
The simulation results are shown in Fig. \ref{NMR}C. We evaluate the performance of primitive and robust pulses by computing their gate fidelities  $F$ over the full system space. The primitive pulse is designed without considering crosstalk (each subsystem is optimized independently), and the robust pulse is generated using the proposed subsystem-optimized control method. 
Although full-system pulse optimization is computationally intractable, evaluating the fidelity of a given pulse remains feasible on a personal computer. As shown, the robust pulses consistently maintain fidelities above (or very close to) 0.99, clearly outperforming the primitive pulses. 
Furthermore, we similarly compare the gate fidelities obtained by repeatedly applying the parallel gates $M$ times; see Fig.~\ref{NMR}D. Both the primitive and robust pulses exhibit exponential fidelity decay, but with markedly different decay rates. For parallel single-qubit gates, the robust pulse reduces the decay rate by roughly a factor of 3 relative to the primitive pulse. For parallel CNOT gates, the reduction is even more pronounced--about a factor of 13--demonstrating the substantial advantage of our robust pulse.
These results confirm that the subsystem-optimized robust control method is capable of producing high-fidelity parallel quantum gates for this 12-qubit NMR system.

\subsection*{Parallel gates in superconducting qubit arrays}

 \begin{figure*}
\includegraphics[width=0.99\linewidth]{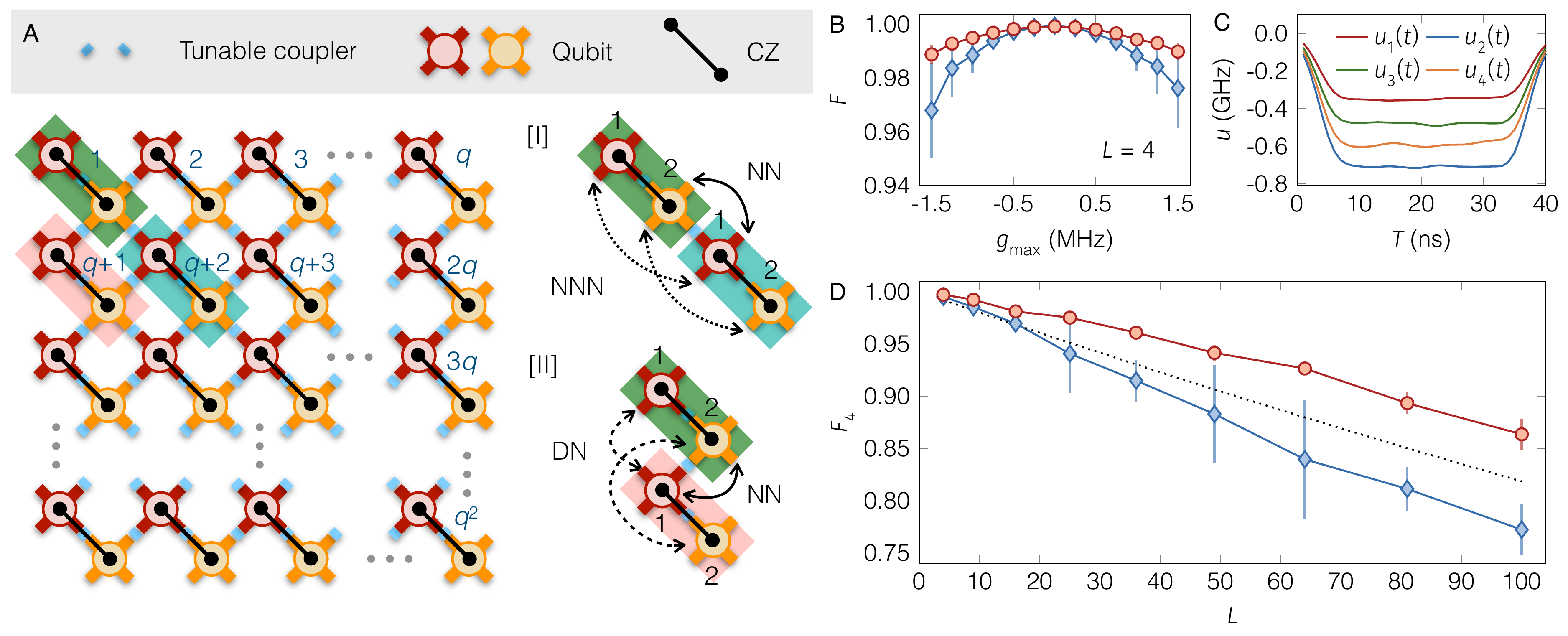}
\caption{\textbf{Implementation of parallel CZ gates on  superconducting qubit arrays.} (\textbf A) Schematic and coupling types of the qubit arrays. The qubit idling frequencies are set to a high band (red, $\omega_{ki}/2\pi \approx 6.15~\mathrm{GHz}$) or a low band (orange, $\omega_{ki}/2\pi \approx 5.85~\mathrm{GHz}$), each with variations within $\pm 0.15~\mathrm{GHz}$.
The anharmonicity is set to $\alpha_{ki}/2\pi\approx -265~\mathrm{MHz}$ with small device-level variations (within $\pm10~\mathrm{MHz}$). The intra-subsystem coupling is fixed at $J_k/2\pi\approx 24~\mathrm{MHz}$ with fluctuations below $\pm2~\mathrm{MHz}$. Nearest-neighbor couplings are sampled within $2|g_{\max}|~\mathrm{MHz}$, while next-nearest and more distant couplings lie within $|g_{\max}|~\mathrm{MHz}$. These parameter ranges follow typical values reported in the literature \cite{google2023suppressing,PhysRevX.11.021058,PhysRevApplied.16.054020,PRXQuantum.3.020301}.
 (\textbf B) Fidelity $F$ of parallel CZ gates as a function of the maximum coupling strength $g_{\max}$ under  primitive (blue lines) and robust (red lines) pulses. The dashed lines indicate the fidelity threshold of 0.99. (\textbf C) Corresponding control waveforms applied to each subsystem for the case $L=4$. (\textbf D) Block fidelity $F_4$ versus the number of CZ gates $L$  under different control pulses, where $g_{\max}=0.8~$MHz. The  dotted reference line corresponds to assumed subsystem fidelities of 0.998.
 For all the cases, the creation and annihilation operators are truncated to the  lowest two energy levels. All error bars show variability across twenty independent runs of random coupling fluctuations. 
  } \label{QubitArray}
\end{figure*} 

We further consider a transmon superconducting qubit array comprising $N$ qubits, exemplified by the Google Sycamore device \cite{arute2019quantum,google2023suppressing}, as illustrated in Fig. \ref{QubitArray}A. Each qubit is coupled to four tunable couplers, allowing for the adjustment of the effective qubit-qubit interactions. Our goal is to implement $L=q^2~(q\in \mathbb{Z}^+)$  high-fidelity controlled-Z (CZ) gates simultaneously, a critical operation with essential applications like quantum error correction \cite{google2023suppressing}.  A primary obstacle is unwanted interactions between qubit pairs implementing the CZ gates, including residual couplings between nearest-neighbor (NN) qubits that cannot be fully turned off, and parasitic couplings between next-nearest-neighbor (NNN) or   diagonal-neighbor (DN) qubits due to their relatively small frequency detuning \cite{PRXQuantum.3.020301}; see Fig. \ref{QubitArray}A. These crosstalk interactions are weak and can be treated as perturbations.

 To apply our method, we first partition the system into $L$ subsystems, each consisting of two qubits that implement a CZ gate; see Fig. \ref{QubitArray}A. In addition, we label these subsystems sequentially from left to right and top to bottom. 
Subsequently, we explicitly express the system Hamiltonian in Eq. (\ref{sysham}) as \cite{krantz2019quantum,PhysRevApplied.10.054062} 
$
H_{S_k}=\sum_{i=1}^2 (\omega_{ki}a^\dag_{ki} a_{ki} + \frac{\alpha_{ki}}{2}a^\dag_{ki}a^\dag_{ki} a_{ki} a_{ki} ) - J_k (a_{k1}-a^\dag_{k1})(a_{k2}-a^\dag_{k2}),
$
and $
  H_{S_k S_j}= -\sum_{i,l=1}^2 g_{kj}^{il} (a_{ki}-a^\dag_{ki})(a_{jl}-a^\dag_{jl}),
$
where $a^\dag_{ki}$ and $a_{ki}$ are the creation and annihilation operators, $\omega_{ki}$ and $\alpha_{ki}$ denote the qubit frequency and anharmonicity of the $i$th qubit in subsystem $k$, respectively, $J_k$ is the coupling strength between the two qubits within the $k$th subsystem, and $g_{kj}^{il}$ represents the coupling strength between the $i$th qubit in subsystem $k$ and the $l$th qubit in subsystem $j$. Due to the negligible interactions between distant subsystems, we focus only on the crosstalk couplings between neighboring subsystems. There are two distinct coupling patterns in this qubit array, as illustrated in Fig. \ref{QubitArray}A. As such, for the $k$th subsystem, the crosstalk terms $H_{S_k S_j}$ to be considered can be reduced to the following cases: (1) if $k~\text{mod}~q=0$ and $k\neq q^2$, then $j = k+q$; (2) if $\lceil \frac{k}{q}\rceil=q$ and $k\neq q^2$, then $j = k+1$; (3) if $k=q^2$, then $j \in \emptyset$; (4) in all other cases, $j \in \{k+1, k+q, k+q+1\}$. Thus, the number of the total crosstalk terms is $3 q^2-4q+1$.
The control pulses adjust the frequency of the first qubit in each subsystem to implement the CZ gates, i.e., $H_{C_k}(t) = \sum_{k=1}^L u_k(t) a^\dag_{k1} a_{k1}, 0\leq t \leq T $.

The simulation results are shown in Figs. \ref{QubitArray}B-D. 
 In Fig. \ref{QubitArray}B, we compare the performance of two pulses by directly simulating the evolution of the entire system involving 4 parallel CZ gates. One is the robust pulse optimized using the proposed method, while the other is a primitive pulse that does not account for crosstalk. The robust pulse achieves a significantly higher fidelity  compared to the primitive pulse, maintaining values above 0.99 for $|g_{\max}| \leq 1$ MHz. Moreover, the robust pulse exhibits  stronger robustness against parameter variations than the primitive pulse, as indicated by its negligible error bars.
 Notably, the robust pulse is optimized assuming a maximum crosstalk strength of $g_{\max} = 1.5$ MHz, yet it remains effective for all  $|g_{\max}| \leq 1.5$ MHz. This eliminates the need to rerun the robust control optimization for different crosstalk strengths. This reveals the key feature of our method: it does not require precise knowledge of the crosstalk amplitude. During optimization, one only needs to specify an upper bound on the crosstalk strength, and the resulting robust control pulse remains effective for all crosstalk values within this bound.
 The control waveforms are shown in Fig. \ref{QubitArray}C, which are sufficiently smooth for experimental implementation.

 To further validate the effectiveness of our method, we apply it to optimize robust pulses for implementing $L = q^2$ parallel CZ gates with $q = 2,\dots,10$. For $L>4$, simulating the full-system evolution on a standard personal computer becomes computationally intractable, preventing direct evaluation of the gate fidelity. Instead, we assess the control performance using a block fidelity $F_4$ obtained by partitioning the entire system into four-subsystem blocks, neglecting the remaining couplings, and evaluating the gate fidelities within these blocks. The simulation results are shown in  Fig. \ref{QubitArray}D. 
  It is clear that the robust pulses consistently produce higher  $F_4$, demonstrating superior resilience to crosstalk. 
It is expected that, if gate fidelities could be evaluated for system sizes larger than four-subsystem blocks, the performance gap between the robust pulse and the primitive pulse would become even more pronounced.
 In the present case,  each subsystem maintains $f_{S_k} > 0.998$, as indicated by the analogous dotted reference line.
 These results highlight the advantage of our subsystem-optimized robust control in mitigating unwanted interactions while realizing large-sized parallel CZ gates.

 \section*{DISCUSSION}
Parallel quantum gates are essential for advancing large-scale quantum information processing, yet their fidelity is often hindered by crosstalk across subsystems. By introducing a subsystem-optimized robust control framework, we show that high-fidelity parallel gates can be achieved by focusing only on limited-size subsystems, rather than the entire system space. This reveals a general principle of subsystem-level robustness: scalable quantum control can be constructed from robust solutions on smaller building blocks. Such a formulation not only drastically reduces computational overhead but also provides a new paradigm for extending control strategies to increasingly large quantum devices. Moreover, our method is independent of precise crosstalk parameters and is applicable to arbitrary qubit connectivity, underscoring its broad applicability.

Building on this foundation, we construct analytical pulses that enable high-fidelity parallel single-qubit gates on coupled NV centers and reduce their noise scaling from exponential to linear. We further demonstrate high-quality parallel multi-qubit gates on both an NMR processor and superconducting arrays, achieving an approximately order-of-magnitude reduction in noise accumulation. These results are particularly significant for near-term quantum computing platforms. By slowing down the rate at which noise accumulates during parallel operations, our method naturally enables the execution of deeper and more complex quantum circuits, thereby extending the computational reach of existing hardware.

Looking forward, the subsystem-optimized robust control framework offers a pathway toward scalable, high-performance quantum technologies. Beyond immediate applications in metrology, simulation, and computation, its integration with tensor network \cite{orus2014practical,berezutskii2025tensor} or Krylov-based algorithms \cite{nandy2025quantum,PhysRevLett.134.030401} could enable efficient exploration of quantum dynamics far beyond the reach of current methods. More broadly, the principle of subsystem-level robustness may serve as a foundation for controlling complex quantum matter, opening new possibilities for building practical quantum technologies and accelerating the transition from proof-of-concept devices to large-scale, fault-tolerant architectures.

\section*{METHODS}
\subsection*{Approximated gate fidelity}
We now explain how the quantum gate fidelity can be simplified for our analysis. We first move to the toggling frame to analyze the crosstalk effects, where the evolution operator is defined as $\tilde U(T)=\mathcal{T}\exp[-i\int_0^T dt\, \tilde H_1(t)]$ with $\tilde H_1(t)=U_{H_0'}^\dagger(t) H_1 U_{H_0'}(t)$. By definition, the full evolution satisfies $U(T)=U_{H_0'}(T)\tilde U(T)$. Expanding $\tilde U(T)$ via the Dyson series gives $U(T)=U_{H_0'}(T)[\mathbb{I}_d - i\int_0^T dt\, U_{H_0'}^\dagger(t)H_1 U_{H_0'}(t) + (-i)^2\int_0^T dt_1 \int_0^{t_1} $ $dt_2\, U_{H_0'}^\dagger(t_1)H_1 U_{H_0'}(t_1) U_{H_0'}^\dagger(t_2)H_1 U_{H_0'}(t_2) + \cdots ]$. To capture the dominant crosstalk contributions, we keep only the first- and second-order perturbative terms, allowing the gate fidelity to be simplified accordingly
\begin{align}\label{tdyson}
	F 
	& = \frac{\left| \operatorname{Tr}(U(T) \overline{U}^\dagger) \right|^2}{ d^2}=\frac{\left| \operatorname{Tr}(\tilde U(T)) \right|^2}{ d^2} \nonumber \\ 
	& \approx \frac{\left| d-\frac{1}{2}  \operatorname{Tr}([\int_0^T dt U_{H_0'}^\dag(t)H_1 U_{H_0'}(t) ]^2) \right|^2}{ d^2}  \nonumber \\
	& \approx 1-\frac{1}{d} \|\mathcal{D}_{U_{H_0'}(T)}(H_1)\|^2 \nonumber \\ 
	& \approx 1-\frac{1}{d} \sum_{k<j}^L \|\mathcal{D}_{U_{H_0'}(T)}(H_{S_k S_j})\|^2, 
\end{align}
where $\|A\|=\sqrt{\text{Tr}(A^\dag A)}$ denotes the Frobenius norm, and we have assumed $U_{H_0'}(T)=\bar U$.

 \subsection*{Reduced directional derivative to subsystem pairs}
 The directional derivative over the full system $f_{S_k S_j} \equiv \|\mathcal{D}_{U_{H_0'}(T)}(H_{S_k S_j})\|^2 / d $ can be reduced  to
\begin{align}
 & \mathcal{D}_{U_{H_0'}(T)}(H_{S_k S_j})
 	= -i U_{H_0'}(T)\int_0^T dt U_{H_0'}^\dag(t)H_{S_k S_j} U_{H_0'}(t)  \nonumber \\ \nonumber
 &=\left\{-i \left[U_{S'_k}(T) \otimes U_{S'_j}(T)\right]\int_0^T dt \left[U^\dag_{S'_k}(t) \otimes U^\dag_{S'_j}(t)\right] \right. \nonumber \\
& \quad \left. \cdot H_{S_k S_j} \left[U_{S'_k}(t) \otimes U_{S'_j}(t)\right] \right\} \otimes \left[\bigotimes^L_{l \ne k,j} U_{S'_l}(T)\right] \nonumber \\ 
 &\equiv \mathcal{D}_{U_{S'_k S'_j}(T)} (H_{S_k S_j})  \otimes U_{S/ \left\{S_k, S_j \right\}}(T), 
\end{align}
where we define $U_{S'_k S'_j}(T)=U_{S'_k}(T) \otimes U_{S'_j}(T)$ and 
$U_{S/ \left\{S_k, S_j \right\}}(T) = \bigotimes^L_{l \ne k,j} U_{S'_l}(T)$. 
As such, we obtain 
\begin{align}
	f_{S_k S_j} & \equiv \|\mathcal{D}_{U_{H_0'}(T)}(H_{S_k S_j})\|^2 / d \nonumber \\
	& = \|\mathcal{D}_{U_{S'_kS'_j}(T)}(H_{S_k S_j})\|^2 / d_{H_{S_k S_j}}.
\end{align}

 \subsection*{Simplified directional derivative for parallel single-qubit quantum gates}
 For parallel single-qubit quantum gates, the norm of the directional deviation $f_{S_k S_j}$  can be expressed as  
\begin{align}
	& f_{S_k S_j}
	 =\frac{1}{4} \left \|\int_0^T U_{S'_kS'_j}^\dag(t)  H_{S_k S_j} U_{S'_kS'_j}(t)  d t \right\|^2 \nonumber  \\
	& =\frac{1}{4} \left \|\int_0^T [U_{S'_k}(t)\otimes U_{S'_j}(t) ]^\dag [g_{kj} \sigma_{z}^k \otimes \sigma_{z}^j] [U_{S'_k}(t)\otimes U_{S'_j}(t) ]  d t \right\|^2  \nonumber\\
	&= \frac{1}{4} \left \| g_{kj} \int_0^T [U_{S'_k}^\dag(t) \sigma_{z}^k  U_{S'_k}(t) ] \otimes [U_{S'_j}^\dag(t) \sigma_{z}^j  U_{S'_j}(t)]     d t \right\|^2 .
\end{align}
Subsequently, we write the Heisenberg-evolved operators as $U_{S'_k}^\dag(t) \sigma_{z}^k  U_{S'_k}(t)=\sum\nolimits_{\mu=x,y,z} r_\mu^k (t)\sigma_\mu^k $ and $U_{S'_j}^\dag(t) \sigma_{z}^j  U_{S'_j}(t)=\sum\nolimits_{v=x,y,z} r_v^j (t)\sigma_v^j $, where $r_\mu^k (t)= \text{Tr}[U_{S'_k}^\dag(t) \sigma_{z}^k  U_{S'_k}(t) \sigma_\mu^k]/2$ and $r_v^j (t)= \text{Tr}[U_{S'_j}^\dag(t) \sigma_{z}^j  U_{S'_j}(t) \sigma_v^j]/2$. Thus, we get
\begin{align}
	f_{S_k S_j}  & =\frac{g_{kj}^2}{4} \left \|\int_0^T \sum\limits_{\mu,v=x,y,z}  r_\mu^k (t)  r_v^j (t) \sigma_\mu^k \otimes   \sigma_v^j d t \right\|^2   \nonumber\\
	&= \frac{g_{kj}^2}{4} \left \| \sum\limits_{\mu,v=x,y,z} \left( \int_0^T r_\mu^k (t)  r_v^j (t) d t\right) \sigma_\mu^k \otimes   \sigma_v^j  \right\|^2, \nonumber \\
	&=  \frac{g_{kj}^2}{4} \sum\limits_{\mu,v=x,y,z} \left( \int_0^T r_\mu^k (t)  r_v^j (t) d t\right)^2,
\end{align}
where we have used the Frobenius norm identity for operators expanded in the Pauli basis, i.e., 
$
	 \|\sum\limits_{\mu,v} a_{\mu v} \sigma_\mu \otimes \sigma_v  \|^2 =\sum_{\mu,v} |a_{\mu v}|^2
$.

\bibliographystyle{naturemag}

\vspace{0.5cm}
\textbf{Acknowledgments:} X.Y. thanks P. Zhao for helpful discussions. 
\textbf{Funding:}
This work is supported by   the Innovation Program for Quantum Science and Technology (2024ZD0300400), the National Natural Science Foundation of China (grant no. 12441502, 12204230), and the Guangdong Provincial Quantum Science Strategic Initiative (GDZX2305006, GDZX2405002).
\textbf{Author contributions:}
 J.L. initiated the project. X.Y. and J.L. conceived the relevant theoretical constructs. X.Y. carried out the simulation and analyzed the data. All authors contributed to discussing the results and writing the manuscript.   
\textbf{Competing interests:}
 The authors declare that there are no competing interests.
\textbf{Data and materials availability:}
All data needed to evaluate the conclusions in the paper are present in the paper and/or the Supplementary Materials.

\end{document}